\documentclass[rapids]{jfm}

\usepackage{graphicx}
\graphicspath{{4_figures/}}
\usepackage{newtxtext}
\usepackage{newtxmath}
\usepackage{natbib}
\usepackage{siunitx}
\usepackage{booktabs}
\usepackage{multirow}
\usepackage{hyperref}
\hypersetup{
    colorlinks = true,
    urlcolor   = blue,
    citecolor  = black,
}

\newcommand{\paragraph}[1]{\medskip\noindent\textbf{#1}\enspace}
\newcommand{\mean}[1]{\langle #1 \rangle}   
\newcommand{\RomanNumeralCaps}[1]
\linenumbers

\usepackage[normalem]{ulem}

\title{Generalised Perturbed Convective Wave Theory}

\author[S. Schoder, E. Bagheri, H. Vincent, T. Brunner]{S. Schoder\aff{1,\corresp{\email{stefan.schoder@tugraz.at}}}
  ,
  E. Bagheri\aff{2},
 H. Vincent\aff{3}
 \and \,\,T. Brunner\aff{1}}

\affiliation{\aff{1}Institute of Fundamentals and Theory in Electrical Engineering, Graz University of Technology, 8010 Graz, Austria
\aff{2}Institute of Fluid Mechanics, Friedrich-Alexander-Universit\"{a}t, 91058 Erlangen, Germany
\aff{3}Ecole Centrale de Lyon, CNRS, Université Lyon 1, INSA Lyon, LMFA, UMR5509, 69130, Ecully, France}

\begin{document}
\maketitle

\begin{abstract}
The theory of the perturbed convective wave equation for compressible flows (cPCWE) is generalised to spatially varying mean-density fields. The resulting equation is an exact scalar reformulation of the acoustic perturbation equations and describes sound generation and propagation in moving inhomogeneous media using a single unknown. The intermediate variables of the associated workflow, in which a Helmholtz decomposition problem, a Poisson equation and the cPCWE are solved successively, are related to the vortical, entropy and acoustic modes of Kov\'asznay, providing a physical interpretation of each processing step. The quantitative accuracy is assessed against fully compressible direct numerical simulations (DNS) of two-dimensional isothermal mixing layer and Lighthill's analogy computed in the same framework at Mach numbers between $M=0.2$ and $M=0.4$, based on the velocity difference across the layer and the ambient speed of sound. Over this range of Mach numbers, the radiated power spans several orders of magnitude. For $M\geq0.25$, the sound power levels obtained using the three methods agree within $0.9\,\mathrm{dB}$, and within $0.5\,\mathrm{dB}$ for $M\geq0.3$. At $M=0.2$, where the acoustic fluctuations are weakest relative to the hydrodynamic ones, Lighthill's analogy over-predicts the radiated power by $2.8\,\mathrm{dB}$. In contrast, the cPCWE deviates from the DNS reference by only $-1.2\,\mathrm{dB}$. This closer agreement is because the cPCWE source term is confined to the vortex-pairing region, while convection and refraction are represented by its convective wave operator. Beyond reproducing the far-field sound, the cPCWE resolves the acoustic field within the shear zone itself, where the DNS' fields are masked by vortical fluctuations.
\end{abstract}


\section{Introduction}
\label{sec:intro}

In free shear flows, only a small fraction of the energy is converted into sound: the acoustic fluctuations are typically several orders of magnitude weaker than the hydrodynamic fluctuations from which they originate. Compressible direct numerical simulation (DNS) resolves them simultaneously and thus provides reference solutions, but it becomes cost-prohibitive at high Reynolds numbers, and the amplitude and length-scale disparities widen as the Mach number decreases \citep{ColoniusLele2004}. Hybrid methods alleviate this difficulty by evaluating the acoustic sources from a flow simulation restricted to the source region, before solving a wave equation for the sound field \citep{Schoder19}.

The wave operators and source terms of these methods largely originate from the acoustic analogy of \citet{lighthill1952_prsl}, in which the compressible Navier-Stokes equations are exactly reformulated as an inhomogeneous wave equation. \citet{Phillips1960} and \citet{Lilley1974} later transferred the effects of mean-flow convection and refraction from the source term to the wave operator, and \citet{Goldstein2003} generalised the approach to non-uniform base flows. In all these formulations, however, acoustic, vortical and entropy contributions remain superimposed within the source region, without any criterion to separate them.

A theoretical basis for such a separation was provided by \citet{Kovasznay53}, who showed that, at first order, small perturbations about a uniform mean flow decompose into independent acoustic, vortical and entropy modes. This result motivated a second family of hybrid methods, in which these contributions are distinguished before the propagation step. In the acoustic perturbation equations of \citet{Ewert2003}, the source terms are filtered to retain only their acoustically effective part, but the perturbation velocity and pressure remain governed by a coupled system. To reduce this coupled system to a single scalar unknown, the perturbed convective wave equation was derived from these equations, together with related formulations based on the convective wave operator of \citet{Pierce1990} \citep{spieser2020sound,schoder2023acoustic}, initially for incompressible flow data. It has recently been extended to compressible flows, leading to the compressible perturbed convective wave equation (cPCWE) \citep{schoder2025perturbed}.%

The formulation, physical interpretation and predictive capabilities of the cPCWE nevertheless remain only partially established. Its original derivation neglects the terms involving mean-density gradients. Moreover, the exact relationship between its intermediate variables and a complete decomposition of the fluctuating field into acoustic, vortical and entropy components, particularly in a non-uniform mean flow, remains to be clarified. Above all, its quantitative accuracy has not yet been assessed against a fully compressible reference solution over 
a wide range of Mach numbers, down to the low values for which the disparity between acoustic and hydrodynamic amplitudes is most severe.

Given the above, the present work pursues three objectives. 
A first objective is to generalise the cPCWE theory to base flows with spatially varying mean density, thereby removing the restriction of its original derivation, and to establish that the resulting scalar equation remains algebraically equivalent to the acoustic perturbation equations within their assumptions.
A second objective is to clarify the physical content of the intermediate variables of the method by relating them to the Kov\'asznay modes, which provides an interpretation of each processing step and explicit recovery relations for the acoustic, vortical and entropy components of the fluctuating field. 
A third objective is to quantify the predictive accuracy of the formulation against a fully compressible reference solution over a range of Mach numbers extending down to values for which the acoustic fluctuations lie several orders of magnitude below the hydrodynamic ones.

The first two objectives are addressed analytically and independently of any particular flow. The third is addressed for forced two-dimensional isothermal mixing layers at Mach numbers $M=\Delta U/c_{\infty}$ between $0.2$ and $0.4$, where $\Delta U$ is the velocity difference across the layer and $c_{\infty}$ the ambient speed of sound.
In these flows, vortex pairings form a tonal, spatially non-compact sound source \citep{Colonius97,BogeyBaillyJuve_aiaaj00}.
The sound radiated by a similar mixing layer was computed by \citet{BogeyBaillyJuve_aiaaj02} from linearised Euler equations forced by an aeroacoustic source term, without prior separation of its acoustic and non-acoustic parts. As the flow is isothermal, this configuration demonstrates the complete workflow and quantifies its accuracy, rather than exercising the mean-density and entropy terms themselves. 
The DNS database of \citet{vincent2023application}, generated over this Mach-number range, supplies both the inputs of the hybrid computations and fully compressible reference solutions, with radiated powers spanning more than three orders of magnitude.
The acoustic fields, directivities and radiated powers predicted by the cPCWE and by a Lighthill computation performed in the same finite-element framework are compared with the DNS results, with particular attention paid to the case with $M=0.2$, in which the radiated power is approximately $150$ times lower than its counterpart at $M=0.3$.

The paper is organised as follows. The extended cPCWE is presented in~\S\ref{sec:theory}. The mixing-layer DNS and the hybrid computations are described in~\S\ref{sec:dns}. Results are presented and discussed in~\S\ref{sec:results}. Concluding remarks are given in~\S\ref{sec:conclusions}.

\section{The perturbed convective wave equation for compressible flows}
\label{sec:theory}

The field variables $(\rho, \boldsymbol{u}, p)$ are first Reynolds-decomposed into a temporal mean component $\mean{(\cdot)}$ and a fluctuating component $(\cdot)'$. The fluctuating velocity component is further decomposed into field components with longitudinal and rotational flow properties, respectively, the irrotational part $\boldsymbol{u}_\mathrm{c}$ and the solenoidal part $\boldsymbol{u}_\mathrm{v}$.
\begin{equation}
  \boldsymbol{u}' = \boldsymbol{u}_\mathrm{v} + \boldsymbol{u}_\mathrm{c},
  \qquad
  \nabla\cdot\boldsymbol{u}_\mathrm{v} = 0,
  \qquad
  \nabla\times\boldsymbol{u}_\mathrm{c} = \boldsymbol{0}\,.
\end{equation}
In the linear limit and according to \citet{Chu1958}, $\boldsymbol{u}_\mathrm{c}\approx\boldsymbol{u}_\mathrm{a}$ is primarily the acoustic field, in which $\boldsymbol{u}_\mathrm{a}$ denotes the acoustic particle velocity. The linear contribution of the fluctuating entropy component is $\boldsymbol{u}_\mathrm{s}^{(1)}=\boldsymbol{0}$. A small second-order entropy contribution $\boldsymbol{u}_\mathrm{s}^{(2)}\sim \frac{s'}{c_\mathrm{p}}\boldsymbol{u}_\mathrm{a}$, may nevertheless be present.
Using the second law of thermodynamics via $\mathrm{d}\rho/\rho = c_0^{-2}\,\mathrm{d}p/\mean{\rho} - \mathrm{d}s/c_\mathrm{p}$, with 
the entropy $s$ and the specific heat at constant pressure $c_\mathrm{p}$, the acoustic perturbation equations \citep{Ewert2003} read
\begin{equation}
  \frac{\partial }{\partial t}\left( \frac{p'}{c_0^2}\right)
  + \mean{\boldsymbol{u}}\cdot\nabla \left( \frac{p'}{c_0^2}\right)
  + \nabla\cdot(\mean{\rho}\,\boldsymbol{u}_\mathrm{a})
  = 
  - \nabla\mean{\rho}\,\cdot\,\boldsymbol{u}_\mathrm{v}
  +\frac{\mean{\rho}}{c_\mathrm{p}}
    \!\left(\frac{\partial s'}{\partial t}
           + \mean{\boldsymbol{u}}\cdot\nabla s'\right) ,
  \label{eq:APE1-p}
\end{equation}
\begin{equation}
  \frac{\partial\boldsymbol{u}_\mathrm{a}}{\partial t}
  + \nabla\!\left(\mean{\boldsymbol{u}}\cdot\boldsymbol{u}_\mathrm{a}\right)
  + \mean{\boldsymbol{\omega}}\times\boldsymbol{u}_\mathrm{a}
  + \nabla\!\frac{p'}{\mean{\rho}}
  = \nabla\Phi_\mathrm{p},
  \label{eq:APE1-u}
\end{equation}
where $c_0^2 = \gamma\mean{p}/\mean{\rho}$ is the square of the isentropic speed of sound, $\gamma$ is the isentropic exponent, $\mean{\boldsymbol{\omega}} = \nabla\times\mean{\boldsymbol{u}}$ is the mean vorticity, the mean-flow convection operator $\mathrm{D}/\mathrm{D}t \equiv \partial/\partial t + \mean{\boldsymbol{u}}\cdot\nabla$, and $\Phi_\mathrm{p}$ is the pseudo-pressure source potential.

The pseudo-pressure source potential $\Phi_\mathrm{p}$ is obtained by solving the following Poisson equation
\begin{equation}
  \nabla^2\Phi_\mathrm{p}
  = -\nabla\cdot\!\left[
      \underbrace{\bigl(\boldsymbol{u}_\mathrm{v}\cdot\nabla\bigr)\boldsymbol{u}_\mathrm{v}
      + \bigl(\mean{\boldsymbol{u}}\cdot\nabla\bigr)\boldsymbol{u}_\mathrm{v}
      + \bigl(\boldsymbol{u}_\mathrm{v}\cdot\nabla\bigr)\mean{\boldsymbol{u}}
      - \left(\frac{\nabla \cdot \boldsymbol{\tau}}{\rho}\right)'}_{-\nabla\Phi_\mathrm{v}}
      + \underbrace{T'\nabla\mean{s}
      - s'\nabla\mean{T}}_{-\nabla\Phi_\mathrm{s}}
      \,
    \right].
  \label{eq:poisson}
\end{equation}
The right-hand side contains three vortex-interaction terms, a viscous-stress ($\boldsymbol{\tau}$) term, and two thermo-acoustic coupling terms. $\Phi_\mathrm{p}$ is a scalar potential completely determined by the solenoidal vortical velocity components, entropy, temperature, and viscous stress. In the incompressible limit, the pressure fluctuation $p_\mathrm{v} =\mean{\rho}\,\Phi_\mathrm{p} $ satisfies the incompressible pressure Poisson equation. The entropy and temperature gradient terms on the right-hand side of \eqref{eq:poisson} are neglected. 


The Helmholtz decomposition \citep{schoder2020postprocessing} of the fluctuating velocity field is
\begin{equation}
  \boldsymbol{u}' = \nabla\times\boldsymbol{A} - \nabla\psi_\mathrm{a}
  \equiv \boldsymbol{u}_\mathrm{v} + \boldsymbol{u}_\mathrm{a},
  \label{eq:helmholtz}
\end{equation}
where $\boldsymbol{A}$ is the vector potential and $\psi_\mathrm{a}$ is the acoustic scalar velocity potential.
The solenoidal vortical velocity $\boldsymbol{u}_\mathrm{v} = \nabla\times\boldsymbol{A}$ carries the entire fluctuating vorticity $\boldsymbol{\omega}' = \nabla\times\boldsymbol{u}' = \nabla\times\boldsymbol{u}_\mathrm{v}$, while $\boldsymbol{u}_\mathrm{a} = -\nabla\psi_\mathrm{a}$ is irrotational and describes the compressible acoustic field. Using the general thermodynamic relation $\mathrm{d}\rho/\rho = c_0^{-2}\,\mathrm{d}p/\mean{\rho} - \mathrm{d}s/c_\mathrm{p}$ \citep{Ewert2003}, the linearised fluctuating density reads $\rho' = p'/c_0^2 - \mean{\rho}\,s'/c_\mathrm{p}$. Substituting into the compressible continuity equation and using $\nabla\cdot\boldsymbol{u}_\mathrm{v}=0$ gives
\begin{equation}
  -\nabla^2\psi_\mathrm{a}
  = \nabla\cdot\boldsymbol{u}'
  = \nabla\cdot\boldsymbol{u}_\mathrm{a}
  = -\frac{1}{\mean{\rho}}\,\frac{\mathrm{D}}{\mathrm{D}t}\left( \frac{p'}{c_0^2}\right)
    + \frac{1}{c_\mathrm{p}}\,\frac{\mathrm{D}s'}{\mathrm{D}t} 
    - \frac{\nabla\mean{\rho}\,\cdot\,\boldsymbol{u}'}{\mean{\rho}}\,.
  \label{eq:helmholtz-poisson}
\end{equation}
Solving \eqref{eq:helmholtz-poisson} yields a scalar potential field $\psi_\mathrm{a}$, from which the vortical velocity fluctuation $\boldsymbol{u}_\mathrm{v} = \boldsymbol{u}' + \nabla\psi_\mathrm{a}$ can be computed. The vortical field is solenoidal by construction $\nabla\cdot\boldsymbol{u}_\mathrm{v} = 0$. The acoustic field is irrotational by construction $\nabla\times\boldsymbol{u}_\mathrm{a} = \boldsymbol{0}$. The kinetic energy of the flow is dominated by the vortical mode $|\boldsymbol{u}_\mathrm{a}|\ll|\boldsymbol{u}_\mathrm{v}|$ at low Mach numbers.  The pseudo-pressure source potential \eqref{eq:poisson} depends only on $\boldsymbol{u}_\mathrm{v}$ and $\mean{\boldsymbol{u}}$, and in the linear approximation of the acoustic field, it does not interact with the source. Essentially, the production and propagation mechanisms are decoupled, allowing a rigorous definition of the energy transfer path.
%
%
With $\boldsymbol{u}_\mathrm{a} = -\nabla\psi_\mathrm{a}$ and applying the Helmholtz decomposition to the mean-vorticity--acoustic-velocity interaction
\begin{equation}
  -\mean{\boldsymbol{\omega}}\times\boldsymbol{u}_\mathrm{a}
  = \mean{\boldsymbol{\omega}}\times\nabla\psi_\mathrm{a}
  = \nabla q_{\omega\times u_\mathrm{a}} + \nabla\times\boldsymbol{B}_{\omega\times u_\mathrm{a}},
\end{equation}
the momentum equation \eqref{eq:APE1-u} yields the fluctuating pressure decomposition using a scalar $q_{\omega\times u_\mathrm{a}}$ and vector $\boldsymbol{B}_{\omega\times u_\mathrm{a}}$ potential
\begin{equation}
  p' = \mean{\rho}\,\frac{\mathrm{D}\psi_\mathrm{a}}{\mathrm{D}t}
       + \mean{\rho}\,q_{\omega\times u_\mathrm{a}}
       + \mean{\rho}\,\Phi_\mathrm{p},
  \label{eq:pressure-decomp}
\end{equation}
identifying three distinct pressure contributions. The first term in \eqref{eq:pressure-decomp} accounts for the acoustic propagation in moving fluids.  The second term arises from the mean-vorticity  and acoustic-velocity interactions. Splitting the last term $\Phi_\mathrm{p} = \Phi_\mathrm{v} + \Phi_\mathrm{s}$ using \eqref{eq:poisson} yields the remaining two contributions: the vortical (hydrodynamic) pressure $p_\mathrm{v} = \mean{\rho}\,\Phi_\mathrm{v}$, and the entropy-induced pressure, $p_\mathrm{s} = \mean{\rho}\,\Phi_\mathrm{s}$.
Substituting \eqref{eq:pressure-decomp} into \eqref{eq:APE1-p} yields the Perturbed Convective Wave Equation for compressible flows 
\begin{equation}
  \frac{\mathrm{D}}{\mathrm{D}t} \left( \frac{\mean{\rho}}{c_0^2} \frac{\mathrm{D}\psi_\mathrm{a}}{\mathrm{D}t}\right)
  + \frac{\mathrm{D}}{\mathrm{D}t}\left( \frac{\mean{\rho}q_{\omega\times u_\mathrm{a}}}{c_0^2}\right)
  - \nabla\,\cdot\, \mean{\rho}\nabla\psi_\mathrm{a}
  = -\frac{\mathrm{D}}{\mathrm{D}t}\left( \frac{\mean{\rho}\Phi_\mathrm{p}}{c_0^2}\right)
  + \frac{\mean{\rho}}{c_\mathrm{p}}\,\frac{\mathrm{D}s'}{\mathrm{D}t}
  - \nabla\mean{\rho}\,\cdot\,\boldsymbol{u}_\mathrm{v}\, .
  \label{eq:cPCWE}
\end{equation}
It extends \citet{schoder2025perturbed} by removing the constraint on the mean density gradients. Apart from its second term, the left-hand side yields Pierce's operator \citep{Pierce1990}. 
The three modes are recovered as follows. Firstly, the acoustic mode is computed by
\begin{equation}
  \boldsymbol{u}_\mathrm{a} = -\nabla\psi_\mathrm{a}\,, \quad
  p_\mathrm{a} = \mean{\rho}\!\left(\frac{\mathrm{D}\psi_\mathrm{a}}{\mathrm{D}t} + q_{\omega\times u_\mathrm{a}}\right)\,, \quad
  \rho_\mathrm{a} = \frac{p_\mathrm{a}}{c_0^2}\,, \quad
  T_\mathrm{a} = \frac{(\gamma-1)\,T_0}{c_0^2}\,\frac{p_\mathrm{a}}{\mean{\rho}}\,, 
  \label{eq:recovery-a}
\end{equation}
with $T_0$ denoting the local mean temperature, the consistent vortical mode is given by
\begin{equation}
\boldsymbol{u}_\mathrm{v} = \boldsymbol{u}' + \nabla\psi_\mathrm{a}\,,\quad
  p_\mathrm{v} = \mean{\rho}\,\Phi_\mathrm{v}\,, \qquad
  \rho_\mathrm{v} = \frac{p_\mathrm{v}}{c_0^2}\,, \qquad
  T_\mathrm{v} = \frac{(\gamma-1)\,T_0}{c_0^2}\,\frac{p_\mathrm{v}}{\mean{\rho}}\,, 
  \label{eq:recovery-v}
\end{equation}
and the entropy mode follows from
\begin{equation}
  \boldsymbol{u}_\mathrm{s} = \boldsymbol{0}\,, \qquad
  p_\mathrm{s} = \mean{\rho}\,\Phi_\mathrm{s}\,, \qquad
  \rho_\mathrm{s} = \frac{p_\mathrm{s}}{c_0^2} - \frac{\mean{\rho}}{c_\mathrm{p}}\,s'\,, \qquad
  T_\mathrm{s} =\frac{(\gamma-1)\,T_0}{c_0^2}\,\frac{p_\mathrm{s}}{\mean{\rho}} + \frac{T_0}{c_\mathrm{p}}\,s' 
  \,.
  \label{eq:recovery-s}
\end{equation}
The decomposition provides a complete separation of independent components.

\section{Mixing-layer DNS and hybrid computations}
\label{sec:dns}

\subsection{Mixing-layer DNS}

The database consists of five forced two-dimensional isothermal mixing layers of air, computed for Mach numbers $M=\Delta U/c_{\infty}=0.20$, $0.25$, $0.30$, $0.35$ and $0.40$, where $\Delta U=U_2-U_1$ is the velocity difference between the two streams. 
The centre velocity is fixed at $U_c=(U_1+U_2)/2=0.5c_{\infty}$, and the Mach number is varied by changing $\Delta U$. 
No splitter plate is included, and the streamwise velocity prescribed at the inlet follows a hyperbolic-tangent profile. 
The corresponding Reynolds number $Re_\omega=\delta_\omega\Delta U/\nu$, based on the inlet vorticity thickness $\delta_\omega$ and the kinematic viscosity $\nu$, ranges from $2000$ to $4000$ over the Mach-number series.
The shear layer is forced at the fundamental instability frequency and its first subharmonic, following \citet{BogeyBaillyJuve_aiaaj00}, to control the vortex-pairing locations.
The combination of this inlet forcing and a downstream sponge zone limits the flow development to the first stage of vortex pairing. 
Consequently, a single well-localised source region contributes predominantly to the radiated field.

The direct numerical simulations are performed using an in-house solver that integrates the unsteady compressible Navier-Stokes equations in Cartesian coordinates $(x,y)$ using high-order explicit finite-difference schemes with low dispersive and dissipative errors. The computational domain extends over $L_x=1200\delta_\omega$ in the streamwise direction and \mbox{$L_y=400\delta_\omega$} in the transverse direction. At the mixing-layer centreline, the transverse mesh spacing is $\Delta y_0=0.1\delta_\omega$ and is progressively increased on both sides of the shear layer. The streamwise spacing is $\Delta x_0=0.2\delta_\omega$ over most of the domain and is progressively increased near the outlet, where a sponge zone is implemented. The time step is $\Delta t=0.9\Delta y_0/c_{\infty}$. Further details are provided by \citet{vincent2023application}.

\subsection{Source-term computation}
\label{sec:methods}

The acoustic field is computed following a hybrid aeroacoustic workflow \citep{Schoder19} in which the DNS provides the velocity, pressure and density fields that serve as input to both the cPCWE and a Lighthill reference computation. The details of the Lighthill reference computation are provided in \citet{schoder2023acoustic}.

The aeroacoustic source terms of the cPCWE are assembled from the DNS fields, following the theory of \S\,\ref{sec:theory}. From the fluctuating velocity $\boldsymbol{u}'$, the vortical component is extracted by solving a Poisson equation \citep{schoder2020postprocessing}. 
The spatial gradients required are approximated on the structured Cartesian DNS grid by fourth-order centred finite differences. The Kronecker tensor product of the corresponding one-dimensional operators forms the two-dimensional gradient operators. The discrete Laplacian is assembled as a sparse Kronecker sum. The sparse system is factorised by sparse LU decomposition. The factors are cached and reused for back-substitution at every time step. For the Helmholtz decomposition of the velocity field, the Poisson equation is singular due to the Neumann boundary conditions and is therefore regularised by Tikhonov regularisation with $\lambda = 10^{-16}$. The vortical velocity is recovered as $\boldsymbol{u}_\mathrm{v} = (\partial A_z/\partial y,\,-\partial A_z/\partial x)$, where the vector potential component in $z$-direction $A_z$ satisfies $\nabla^2 A_z = -\omega_z$, with $\omega_z$ being the vorticity component in $z$-direction. The entropy and mean density gradient contributions to the decomposition are neglected as a result of the thermodynamic state and the low streamwise variation of the mean density at the considered Mach numbers. Homogeneous Neumann conditions are applied at the boundaries of the computational domain.

The pseudo-pressure source potential $\Phi_\mathrm{p}$ is subsequently
obtained from the vortex-interaction part of the Poisson equation
\eqref{eq:poisson},
\begin{equation}
  \nabla^2\Phi_\mathrm{p}
  = -\nabla\!\cdot\!\bigl[
       (\boldsymbol{u}_\mathrm{v}\!\cdot\!\nabla)\boldsymbol{u}_\mathrm{v}
     + (\mean{\boldsymbol{u}}\!\cdot\!\nabla)\boldsymbol{u}_\mathrm{v}
     + (\boldsymbol{u}_\mathrm{v}\!\cdot\!\nabla)\mean{\boldsymbol{u}}\bigr],
  \label{eq:poisson-solved}
\end{equation}
where $\mean{\boldsymbol{u}}$ is the time average of the DNS velocity field.
The viscous stress contributions to \eqref{eq:poisson}
are verified to be more than two orders of magnitude smaller than the three
retained vortex-interaction terms and are accordingly neglected.
The entropy terms are neglected as a result of the thermodynamic state.
The pseudo-pressure Poisson equation is solved with homogeneous Dirichlet conditions
$\Phi_\mathrm{p} = 0 \, \mathrm{m^2/s^2}$ 
\citep{schoder2024aeroacoustic}.
The mean-flow convective derivative is computed by using second-order central finite differences for the time derivative, 
while the convective contribution uses the fourth-order
gradient operator, with the temporal mean velocity field $\mean{\boldsymbol{u}}$ as the advecting velocity. The
assembled source time series is prescribed pointwise as a forcing on the
right-hand side of the acoustic finite-element equations at each time step.
To prevent the impulsive response of the acoustic field to the abrupt source
onset from contaminating the statistics, the source is ramped up through a
smooth blending window over the first $N_\mathrm{trans} = 70$ time steps.
The assembled source time series is then transferred from the DNS grid to the acoustic mesh by conservative supermesh interpolation. 
Each source value is distributed to the FEM nodes, while conserving the field energy. These preprocessing steps are implemented in pyCFS \citep{wurzinger2024pycfs}.
\vspace{-6mm}
\subsection{Acoustic simulation}
\label{sec:acoustic-sim}

The acoustic simulations are performed using the open-source finite-element framework openCFS \citep{CFS}. 
The acoustic potential $\psi_\mathrm{a}$ is computed on a structured 
mesh with uniform element size $h_\mathrm{ac} = \Delta x = \Delta y =
5\times10^{-4}\,\mathrm{m}$. About 27 elements resolve the acoustic wavelength at the dominant vortex-pairing frequency. Under
the isothermally initiated and weakly inhomogeneous conditions of the present mixing layer,
the entropy and $\nabla\mean{\rho}\!\cdot\!\boldsymbol{u}_\mathrm{v}$
contributions on the right-hand side of \eqref{eq:cPCWE} are negligible. Furthermore, $q_{\omega\times u_\mathrm{a}}$ is set to zero. The
equation to be solved therefore reduces to
\begin{equation}
  \frac{\mathrm{D}}{\mathrm{D}t}\!\left(\frac{\mean{\rho}}{c_0^{2}}\,\frac{\mathrm{D}\psi_\mathrm{a}}{\mathrm{D}t}\right)
  - \nabla\!\cdot\!\bigl(\mean{\rho}\nabla\psi_\mathrm{a}\bigr)
  = -\frac{\mathrm{D}}{\mathrm{D}t}\!\left(\frac{\mean{\rho}\,\Phi_\mathrm{p}}{c_0^{2}}\right),
  \label{eq:cPCWE-solved}
\end{equation}
with the values for $c_0$ and $\mean{\rho}$ computed from the DNS fields.
Equation \eqref{eq:cPCWE-solved} is discretised in space by bilinear quadrilateral finite elements, with
source integrals evaluated by sixth-order Gauss--Legendre quadrature. 
Temporal integration uses the Hilber--Hughes--Taylor $\alpha$-method
(HHT-$\alpha$)~\citep{hht1977}, configured with $\alpha = -0.3$.%
The time step used in the acoustic computations corresponds to eight DNS time steps.
Each computation comprises $N_t=423$ time steps, spanning ten vortex-pairing periods. 
The acoustic meshes contain $406\,785$ nodes at $M = 0.2$ and $267\,430$ nodes
at $M = 0.4$. Non-reflecting outer boundaries are
enforced by perfectly matched layers consisting of five elements in the layer thickness direction. The acoustic pressure is recovered
\textit{a posteriori} from \eqref{eq:recovery-a} 
and the acoustic particle velocity from $\boldsymbol{u}_\mathrm{a} = -\nabla\psi_\mathrm{a}$.

\section{Results}
\label{sec:results}

\subsection{Decomposition of the fluctuating field}
\label{sec:decomposition}

\begin{figure}
  \centerline{\includegraphics[width=0.9\linewidth]{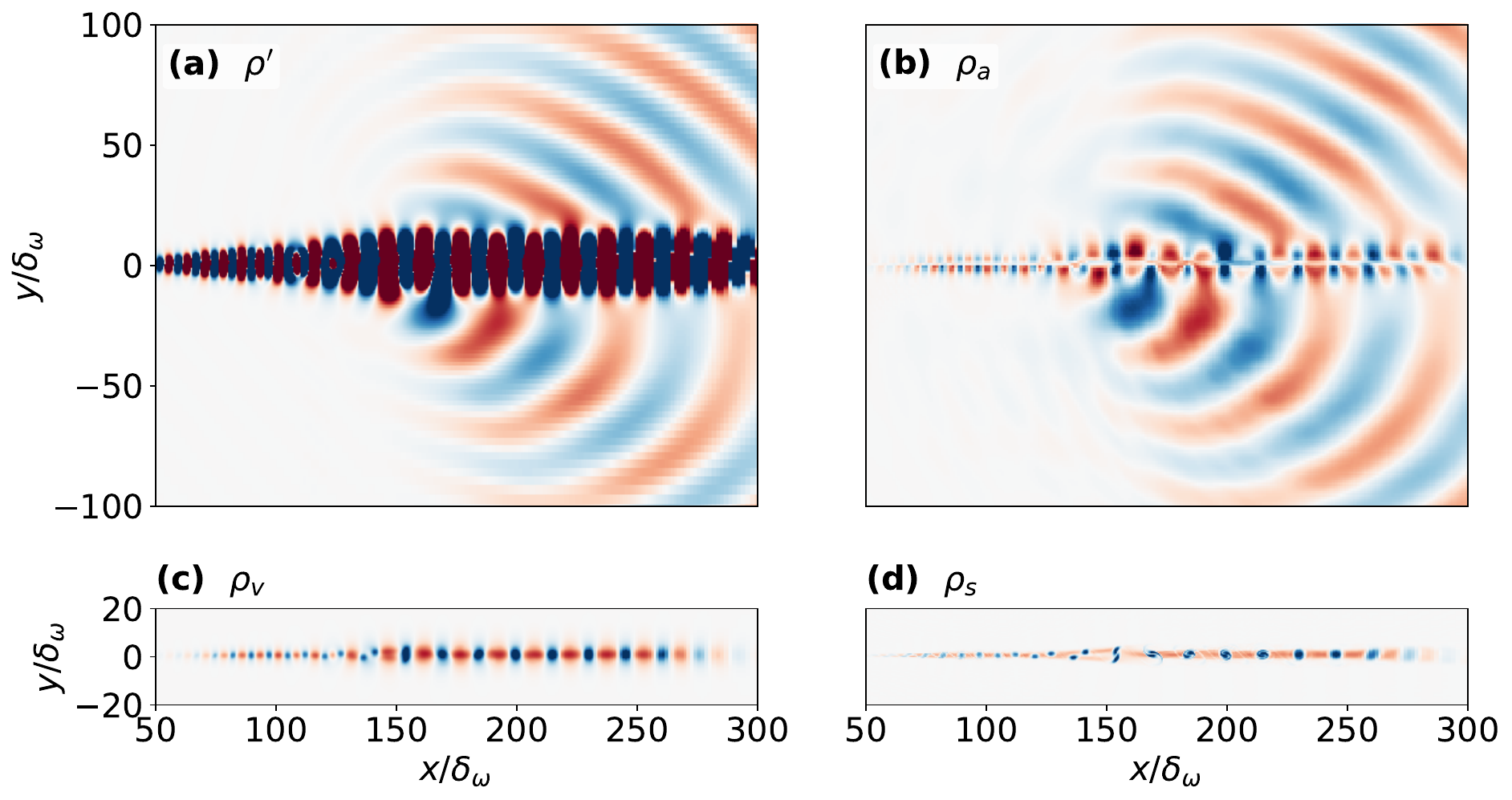}}
\caption{Density fluctuations for $M=0.30$:
(a) DNS field $\rho'$; (b) acoustic component $\rho_\mathrm{a}$; (c) vortical component $\rho_\mathrm{v}$ and (d) entropy component
$\rho_\mathrm{s}$. The colour
scales span $\pm 1.7\times10^{-4}\,\mathrm{kg\,m^{-3}}$ in (a,b),
$\pm 2.85\times10^{-2}\,\mathrm{kg\,m^{-3}}$ in (c) and
$\pm 5.9\times10^{-4}\,\mathrm{kg\,m^{-3}}$ in (d), from blue to red.}
\vspace{-4.3mm}
  \label{fig:decomposition}
\end{figure}

The decomposition of the density fluctuations obtained from the cPCWE is illustrated in Figure~\ref{fig:decomposition} for $M=0.3$.
A snapshot of the DNS density fluctuation $\rho'$ is compared with the acoustic, vortical and entropy contributions $\rho_\mathrm{a}$, $\rho_\mathrm{v}$ and $\rho_\mathrm{s}$, determined from
\eqref{eq:recovery-a}, \eqref{eq:recovery-v} and \eqref{eq:recovery-s}, respectively.
The DNS field is dominated, in the shear
zone, by fluctuations of alternating sign associated with the Kelvin--Helmholtz
vortices forming, convecting and pairing downstream. Their amplitude is highest
around the vortex-pairing location, near $x=150\,\delta_\omega$ for this Mach
number, and decreases rapidly on both sides of the mixing layer. Outside the
layer, circular acoustic wavefronts originating from the pairing region propagate
in the two streams.
In figure~\ref{fig:decomposition}(b), the acoustic component closely reproduces these wavefronts outside
the mixing layer. More importantly, it isolates and reveals the acoustic field within and near the shear zone, where the vortical fluctuations in the DNS results mask it. The amplitude of the acoustic density fluctuations is more than two orders of magnitude lower than that of the vortical fluctuations observed in figure~\ref{fig:decomposition}(c).

The vortical component contains the
energetic structures of the mixing layer and vanishes away from the shear zone,
consistently with the fact that it carries the entire fluctuating vorticity. The
entropy component is also confined to
the mixing layer, with an amplitude nearly two orders of magnitude lower than
that of the vortical fluctuations, as expected for an isothermal
mixing layer. 
The three components therefore differ by up to two orders of magnitude in amplitude and length scale, although they all fluctuate at
the same frequencies, imposed by the harmonic forcing of the shear layer.
Consequently, they can not be separated by temporal or spectral filtering. By contrast, the decomposition provided by the cPCWE isolates them throughout the field, both inside and outside the source region, and provides a rigorous framework for identifying their individual contributions to sound generation and propagation mechanisms.

%

\subsection{Acoustic directivity, radiated power and Mach-number scaling}
\label{sec:energy_results}
\label{sec:scaling_results}

\begin{figure}
  \begin{center}
    \begin{minipage}{0.43\linewidth}
      \centerline{\small (a)}
      \vspace{2mm}
      \centerline{\includegraphics[width=\linewidth]
      {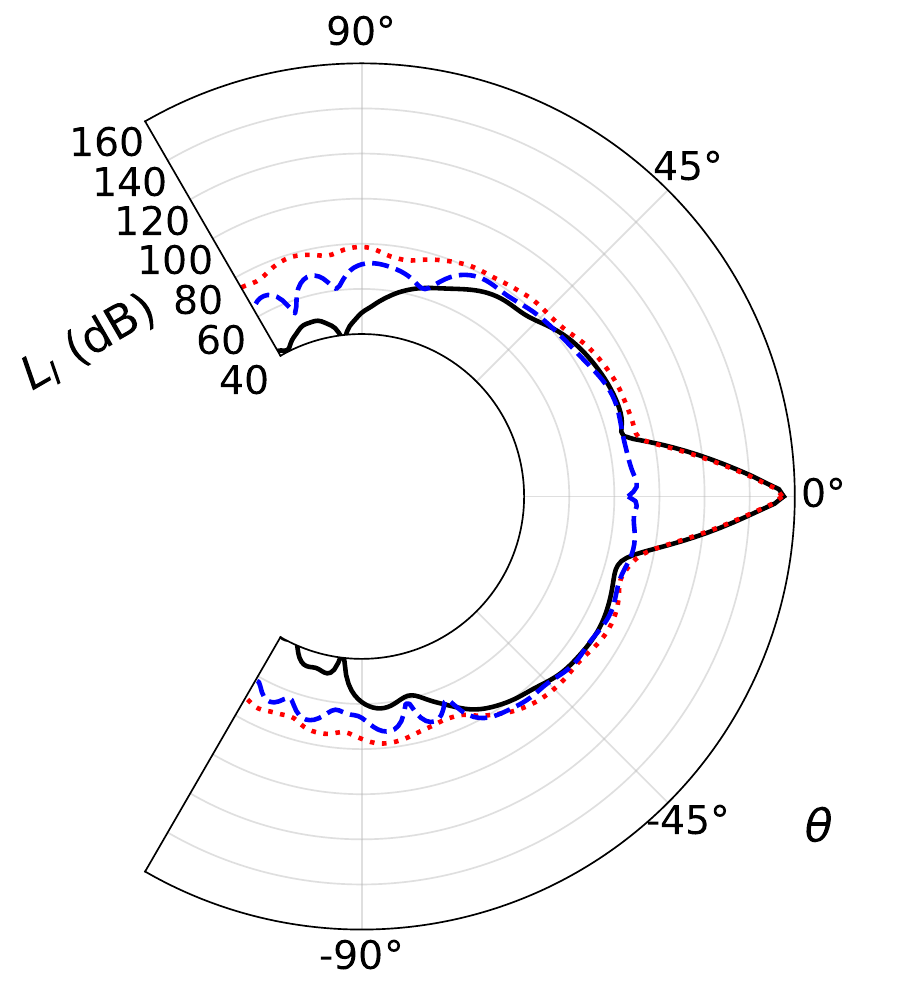}}
    \end{minipage}
    \hfill
    \begin{minipage}{0.43\linewidth}
      \centerline{\small (b)}
      \vspace{2mm}
      \centerline{\includegraphics[width=\linewidth]
      {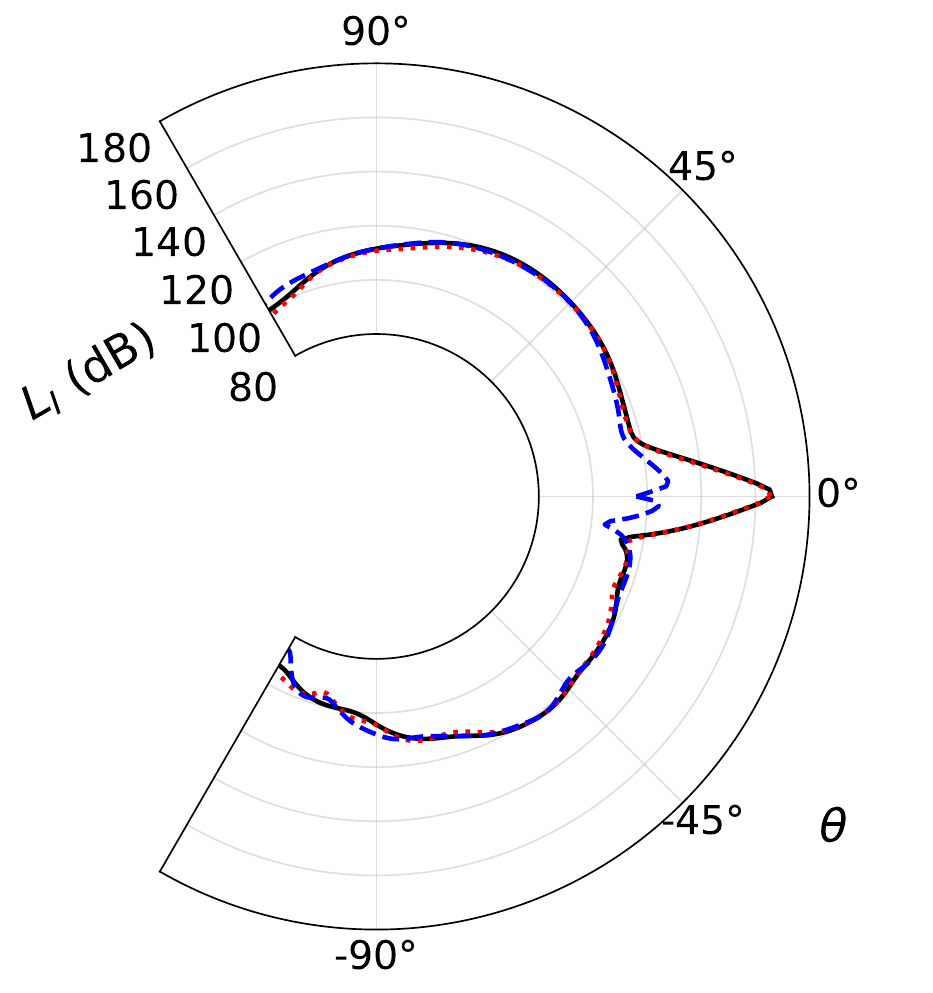}}
    \end{minipage}
  \end{center}
  \caption{Directivity of the acoustic intensity level $L_I$ on the arc
$r = 90\,\delta_\omega$ centred at the pairing location for
(a) $M=0.2$ and (b) $M=0.4$, with --- DNS;
\textcolor{blue}{- -~-} cPCWE; and
\textcolor{red}{$\cdots\cdots$} results based on Lighthill's theory.}
  \label{fig:directivity_all}
\end{figure}

The directivity of the sound field is examined in
figure~\ref{fig:directivity_all}, where the levels
$L_I=10\log_{10}(I/I_\mathrm{ref})$, with $I_\mathrm{ref}=10^{-12}\,\mathrm{W\,m^{-2}}$,
of the radial component of the acoustic intensity
$I(\theta)=p'^{2}_{\mathrm{rms}}(\theta)/(\mean{\rho}\,c_0)$ are represented as a
function of the polar angle $\theta$.
The levels are computed on the arc of radius
$r=90\,\delta_\omega$ centred at the vortex-pairing location, using the fluctuating pressure fields from the DNS, Lighthill computations, and the acoustic pressure $p_\mathrm{a}$ predicted by the cPCWE. All methods reproduce the characteristic radiation pattern of a subsonic mixing layer. Two lobes above and below the layer are present, with maximum radiation at shallow angles from the downstream axis and levels decreasing towards the sideline and upstream directions. The narrow peak within $|\theta| < 10^\circ$ corresponds to the hydrodynamic near field convected along the layer axis for the DNS and Lighthill's theory results and is excluded from the sound power integration. For the cPCWE results, there is no hydrodynamic near field visible. At $M=0.4$, the three predictions are
nearly indistinguishable at all angles except for $|\theta| < 10^\circ$. At $M=0.2$, the intensity levels span
more than $120\,\mathrm{dB}$ between the near-axis peak and the quietest upstream
angles. In that case, the levels given by Lighthill's analogy lie several
decibels above the DNS levels over the sideline and upstream sectors, whereas the
cPCWE results remain close to the DNS levels at all angles. 

\begin{figure}
  \begin{center}\includegraphics[width=0.56\linewidth]{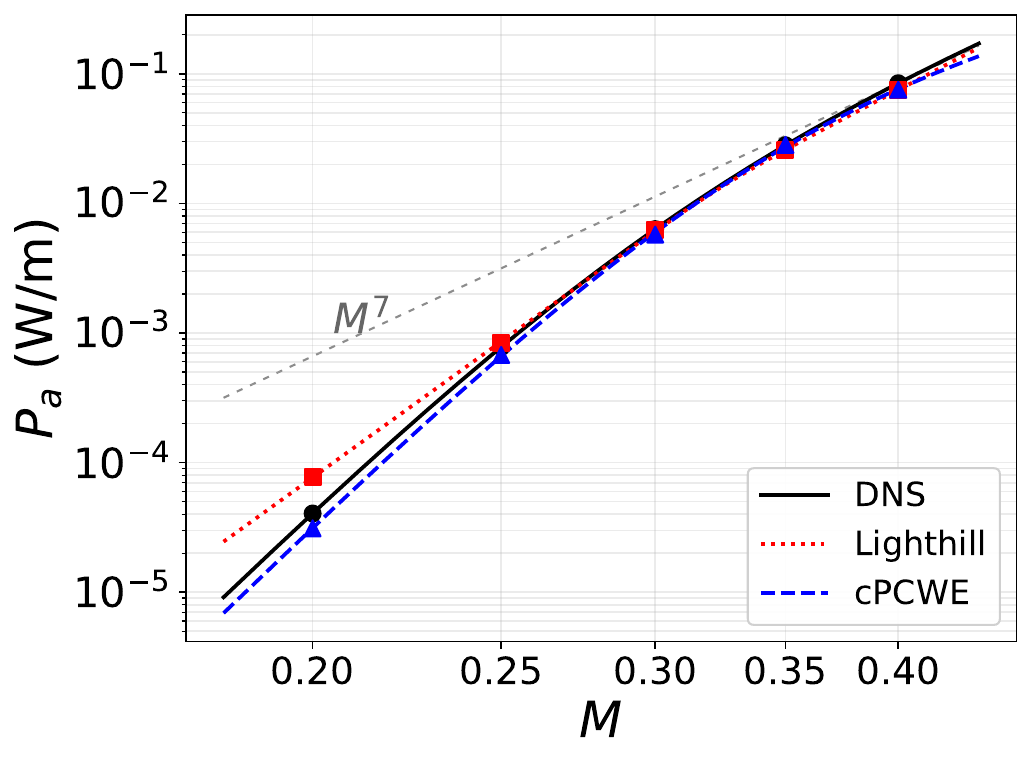}
  \end{center}
  \caption{Radiated acoustic power per unit span $P_a$ as a function of the Mach
number.} 
  \label{fig:mach}
\end{figure}

The sound power per unit span $P_a$ radiated by the mixing layers is then
obtained by integrating the acoustic intensity over the arc $r=90\,\delta_\omega$,
for $\theta\in[-170^\circ,-10^\circ]\cup[+10^\circ,+170^\circ]$ in order to
exclude the hydrodynamic near field. It is represented in figure~\ref{fig:mach}
as a function of the Mach number, together with the $M^7$ power law expected for
compact quadrupole sources in two dimensions \citep{FfowcsWilliams1969}. For the
three methods, the radiated power increases by more than three orders of
magnitude between $M=0.2$ and $M=0.4$. It grows more steeply than the $M^7$
law, with an effective slope decreasing as the Mach number increases and
approaching this law at the upper end of the range, in agreement with the DNS
results of \citet{vincent2023application} obtained for the same flows. The cPCWE
and Lighthill predictions are nearly superimposed on the DNS results at all Mach
numbers, except at $M=0.2$, where the power given by Lighthill's analogy
visibly departs from the DNS value.

The sound power levels $L_\mathrm{W}$ (ref. $10^{-12}\,\mathrm{W/m}$, in dB) obtained at the five Mach numbers are
collected in table~\ref{tab:energy}, together with the differences
$\Delta L_\mathrm{W}$ relative to the DNS. For $M\geq0.25$, the three methods
agree within $0.9\,\mathrm{dB}$, and within $0.5\,\mathrm{dB}$ for $M\geq0.3$.
The DNS levels are moreover slightly higher than or equal to the cPCWE
predictions, by at most $0.6\,\mathrm{dB}$ over this range, which can be
attributed to residual hydrodynamic pressure sampled on the portions of the arc
closest to the shear layer.
The mixing layer at $M=0.2$ is the most demanding case: its radiated power is
roughly $150$ times lower than at $M=0.3$, and the intensity levels to be
predicted on the arc span more than $120\,\mathrm{dB}$, from about
$160\,\mathrm{dB}$ near the layer axis down to $40\,\mathrm{dB}$ in the upstream
direction. In that case, the power level provided by Lighthill's analogy exceeds
the DNS value by $+2.8\,\mathrm{dB}$. Its source term retains mean-flow
convection and refraction, and hence does not vanish in the radiation zone, where the acoustic field results from a superposition of the source and the wave operator. In the cPCWE, the convection and refraction effects are carried by the convective wave operator and the source is confined to the pairing region, decoupling the generation from the propagation effects. The cPCWE deviates from the DNS value by only $-1.2\,\mathrm{dB}$.
For all the Mach numbers considered, the cPCWE thus
predicts the radiated sound power within $1.2\,\mathrm{dB}$ of the fully
compressible reference.

\begin{table}
  \centering
  \caption{Sound power level $L_\mathrm{W}$ 
           and the difference relative to DNS $\Delta L_\mathrm{W}$.}
  \label{tab:energy}
  \setlength{\tabcolsep}{6pt}
  \begin{tabular}{@{}l l c cc cc@{}}
    \toprule
    \multirow{2}{*}{Case} & & DNS & \multicolumn{2}{c}{Lighthill} & \multicolumn{2}{c}{cPCWE} \\
    \cmidrule(lr){4-5} \cmidrule(lr){6-7}
    $M$ & $Re_\omega$ & $L_\mathrm{W}$\,(dB) & $L_\mathrm{W}$\,(dB) & $\Delta L_\mathrm{W}$\,(dB) & $L_\mathrm{W}$\,(dB) & $\Delta L_\mathrm{W}$\,(dB) \\
    \midrule
    $0.20$ & 2000 & $76.1$ & $78.9$ & $+2.8$ & $74.9$ & $-1.2$ \\
    $0.25$ & 2500 & $88.9$ & $89.2$ & $+0.3$ & $88.3$ & $-0.6$ \\
    $0.30$ & 3000 & $98.0$ & $98.0$ & $\pm0.0$ & $97.6$ & $-0.4$ \\
    $0.35$ & 3500 & $104.5$ & $104.1$ & $-0.4$ & $104.5$ & $\pm0.0$ \\
    $0.40$ & 4000 & $109.3$ & $108.8$ & $-0.5$ & $108.8$ & $-0.5$ \\
    \bottomrule
  \end{tabular}
\end{table}

\section{Conclusions}
\label{sec:conclusions}

The perturbed convective wave equation for compressible flows (cPCWE) has been generalised to spatially varying mean-density fields, making it an exact scalar reformulation of the acoustic perturbation equations. The cPCWE predictions were assessed against fully compressible DNS of forced two-dimensional mixing layers at Mach numbers between $0.2$ and $0.4$, and were compared with predictions obtained from the Lighthill equation. The cPCWE has been found to separate the fluctuating
field into its three Kov\'asznay components, inside the shear zone where the acoustic fluctuations are masked by
the vortical ones, and to predict the radiated sound field in agreement with
the DNS over the whole Mach-number range. 
At the lowest Mach number, the predictions of the cPCWE remain accurate while those of Lighthill's analogy deteriorate, because the cPCWE source term is confined to the vortex-pairing region, leaving the radiation zone free of source contributions.

By combining an exact scalar formulation with a physically interpretable
decomposition, the cPCWE thus provides a robust alternative to acoustic
analogies for flows in which the acoustic fluctuations are orders of magnitude
weaker than the hydrodynamic ones. In future studies, it will be interesting
to apply the method to three-dimensional turbulent and non-isothermal flows,
for which the mean-density and entropy terms retained in the present
formulation are expected to come into play.

\backsection[Declaration of interests]{The authors report no conflict of interest.}

\vspace{-10pt}
\bibliographystyle{jfm}
\bibliography{jfm}

@article{hht1977,
  author  = {Hilber, H. M. and Hughes, T. J. R. and Taylor, R. L.},
  title   = {Improved numerical dissipation for time integration algorithms in structural dynamics},
  journal = {Earthquake Engng Struct. Dyn.},
  volume  = {5},
  number  = {3},
  pages   = {283--292},
  year    = {1977},
  doi     = {10.1002/eqe.4290050306},
}

@article{lighthill1952_prsl,
	author = {Lighthill, M. J.},
	title = {On Sound Generated Aerodynamically. {I}. {G}eneral Theory},
	volume = {211},
	number = {1107},
	pages = {564--587},
	year = {1952},
	doi = {10.1098/rspa.1952.0060},
	journal = {Proc. R. Soc. Lond. A},
}

@article{schoder2024aeroacoustic,
  title={Aeroacoustic source potential based on {P}oisson’s equation},
  author={Schoder, S. and Bagheri, E. and Spieser, {\'E}.},
  journal={AIAA J.},
  volume={62},
  number={7},
  pages={2772--2782},
  year={2024},
  publisher={American Institute of Aeronautics and Astronautics},
  doi={10.2514/1.J063792}
}

@misc{wurzinger2024pycfs,
  title={{pyCFS}-data: data processing framework in {P}ython for {openCFS}},
  author={Wurzinger, A. and Heidegger, P. and Schoder, S.},
  howpublished={arXiv:2405.03437},
  year={2024},
  doi={10.48550/arXiv.2405.03437}
}

@article{schoder2020postprocessing,
  title={Postprocessing of direct aeroacoustic simulations using {H}elmholtz decomposition},
  author={Schoder, S. and Roppert, K. and Kaltenbacher, M.},
  journal={AIAA J.},
  volume={58},
  number={7},
  pages={3019--3027},
  year={2020},
  publisher={American Institute of Aeronautics and Astronautics},
  doi = {10.2514/1.J058836}
}

@misc{CFS,
  title={{openCFS}: Open source finite element software for coupled field simulation--part acoustics},
  author={Schoder, S. and Roppert, K.},
  howpublished={arXiv:2207.04443},
  year={2022},
  doi = {10.48550/arXiv.2207.04443}
}

@Article{Chu1958,
  author    = {Chu, B. and Kov{\'a}sznay, L. S. G.},
  journal   = {J. Fluid Mech.},
  title     = {Non-linear interactions in a viscous heat-conducting compressible gas},
  year      = {1958},
  number    = {5},
  pages     = {494--514},
  volume    = {3},
  doi       = {10.1017/S0022112058000148},
  publisher = {Cambridge Univ Press},
}

@Article{Colonius97,
  author  = {T. Colonius and S.~K. Lele and P. Moin},
  title   = {Sound generation in a mixing layer},
  journal = {J. Fluid Mech.},
  year    = {1997},
  volume  = {330},
  pages   = {375--409}
}

@Article{Phillips1960,
  author  = {O.~M.~Phillips},
  journal = {J. Fluid Mech.},
  title   = {On the generation of sound by supersonic turbulent shear layers},
  year    = {1960},
  pages   = {1--18},
  volume  = {9},
  doi     = {10.1017/S0022112060000888},
}

@TechReport{Lilley1974,
  author = {G.~M.~Lilley},
  title  = {On the noise from jets},
  year   = {1974},
  number = {CP-131},
  institution = {AGARD}
}

@Article{Kovasznay53,
  author  = {L.~S.~G. Kov{\'a}sznay},
  title   = {Turbulence in supersonic flow},
  journal = {J. Aeronaut. Sci.},
  year    = {1953},
  volume  = {20},
  pages   = {657--674}
}

@Article{Schoder19,
  author  = {S. Schoder and M. Kaltenbacher},
  title   = {Hybrid aeroacoustic computations: state of art and new achievements},
  journal = {J. Theoret. Comput. Acoust.},
  year    = {2019},
  volume  = {27},
  pages   = {1950020}
}

@Article{Pierce1990,
  author  = {A.~D. Pierce},
  title   = {Wave equation for sound in fluids with unsteady inhomogeneous flow},
  journal = {J. Acoust. Soc. Am.},
  year    = {1990},
  volume  = {87},
  pages   = {2292--2299}
}

@article{vincent2023application,
  title={Application of the complex differentiation method to the sensitivity analysis of aerodynamic noise},
  author={Vincent, H. and Bogey, C.},
  journal={Comput. Fluids},
  volume={264},
  pages={105965},
  year={2023},
  publisher={Elsevier},
doi = {10.1016/j.compfluid.2023.105965}
}

@article{schoder2023acoustic,
  title={Acoustic modeling using the aeroacoustic wave equation based on {P}ierce’s operator},
  author={Schoder, S. and Spieser, {\'E}. and Vincent, H. and Bogey, C. and Bailly, C.},
  journal={AIAA J.},
  volume={61},
  number={9},
  pages={4008--4017},
  year={2023},
  publisher={American Institute of Aeronautics and Astronautics},
  doi={10.2514/1.J062558}
}

@article{schoder2025perturbed,
  title={Perturbed convective wave equation for low-to-medium {M}ach number subsonic flows},
  author={Schoder, S. and Bagheri, E. and Bogey, C. and Bailly, C.},
  journal={J. Sound Vib.},
  volume={623},
  pages={119549},
  year={2025},
  publisher={Elsevier}
}

@article{spieser2020sound,
  title={Sound propagation using an adjoint-based method},
  author={Spieser, {\'E}. and Bailly, C.},
  journal={J. Fluid Mech.},
  volume={900},
  pages={A5},
  year={2020},
  publisher={Cambridge University Press},
doi = {10.1017/jfm.2020.469}
}

@article{Ewert2003,
    title = {Acoustic perturbation equations based on flow decomposition via source filtering},
    volume = {188},
    doi = {10.1016/S0021-9991(03)00168-2},
    number = {2},
    journal = {J. Comput. Phys.},
    publisher = {Elsevier},
    author = {Ewert, R. and Schröder, W.},
    year = {2003},
    pages = {365--398},
}

@article{BogeyBaillyJuve_aiaaj00,
	author = {Bogey, C. and Bailly, C. and Juv{\'e}, D.},
	title = {Numerical simulation of sound generated by vortex pairing in a mixing layer},
	journal = {AIAA J.},
	volume = {38},
	number = {12},
	pages = {2210--2218},
	year = {2000},
	doi = {10.2514/2.906}
}

@article{BogeyBaillyJuve_aiaaj02,
  title={Computation of flow noise using source terms in linearized {E}uler's equations},
  author={Bogey, C. and Bailly, C. and Juv{\'e}, D.},
  journal={AIAA J.},
  volume={40},
  number={2},
  pages={235--243},
  year={2002},
  doi={10.2514/2.1665}
}

@article{FfowcsWilliams1969,
  title={Hydrodynamic noise},
  author={Ffowcs Williams, J. E.},
  journal={Annu. Rev. Fluid Mech.},
  volume={1},
  pages={197--222},
  year={1969},
  doi={10.1146/annurev.fl.01.010169.001213}
}

@article{Goldstein2003,
  title={A generalized acoustic analogy},
  author={Goldstein, M. E.},
  journal={J. Fluid Mech.},
  volume={488},
  pages={315--333},
  year={2003},
  doi={10.1017/S0022112003004890}
}

@article{ColoniusLele2004,
  title={Computational aeroacoustics: progress on nonlinear problems of sound generation},
  author={Colonius, T. and Lele, S. K.},
  journal={Prog. Aerosp. Sci.},
  volume={40},
  number={6},
  pages={345--416},
  year={2004},
  doi={10.1016/j.paerosci.2004.09.001}
}

\end{document}